\documentclass[superscriptaddress,aip,amsmath,amssymb,floatfix,reprint,raggedbottom]{revtex4-1}
\usepackage[pdftex]{graphicx}
\usepackage{amsmath}
\usepackage{lipsum}
\usepackage{color}
\usepackage{soul}
\usepackage{balance}
\usepackage{fancyhdr}
\usepackage{booktabs}
\usepackage{enumitem}
\usepackage{listings}
\usepackage{hyperref}

\hypersetup{colorlinks=true,linkcolor=blue,citecolor = blue,urlcolor = blue, linktocpage}

\fancypagestyle{firststyle}
{
\fancyhf{}

\pagestyle{fancy}
}

\usepackage{fancyhdr}

\begin{document}
%\linenumbers

\title{Can Negative Capacitance Induce Superconductivity?}
\author{Supriyo Datta}
\affiliation{Elmore Family School of Electrical and Computer Engineering, Purdue University, West Lafayette, IN, 47906 USA}

%\author{...}
%\affiliation{Department of Electrical and Computer Engineering, Purdue University, West Lafayette, IN, 47906 USA}

%\author{et al.}
%\affiliation{??}
\date{\today}

\begin{abstract}
Superconductivity was originally observed in 3D metals caused by an effective attraction between electrons mediated by the electron-phonon interaction.  Since then there has been a lot of work on 2D conductors including the possibility of alternative mechanisms that can lead to an effective attractive interaction. Inspired by the experimental demonstration of both steady-state and transient negative capacitance in a variety of structures, this paper investigates the possibility of superconductivity in a two-dimensional conductor embedded in a \textit{negative permittivity medium} whose role is to turn the normally repulsive Coulomb interaction into an attractive one. A weak coupling BCS theory is used to identify the key parameters that have to be optimized to observe a superconducting transition, especially the need for a small effective negative permittivity, which could be obtained by balancing a negative permittivity medium with a positive permittivity one.

\end{abstract}
\pacs{}
\maketitle
\thispagestyle{firststyle}

\section{Introduction}
\vspace{-0.1in}

Superconductivity is an intriguing phenomenon that is believed to arise in a wide variety of many-fermion systems due to two-body attractive interactions which introduce a very special kind of correlation among the one-fermion states. This state is commonly described by the Bardeen-Cooper-Schrieffer (BCS) theory [BCS 1957, PWA 1958, MLC 1964, PBA 1982] which was originally advanced to describe the superconducting state in 3D metals caused by an effective attraction between electrons mediated by the electron-phonon interaction. Since then there has been a lot of work on 2D superconductors which introduce many subtle issues summarized in a recent review [SAI 2021]. The recent discovery of superconductivity in twisted bilayer graphene has led to increased interest in 2D superconductivity including the possibility of alternative mechanisms that can lead to an effective attractive interaction, see for example, [BLG 2018, CEA 2021].

An enhancement in the superconducting transition temperature by $\sim 0.15K$  was observed experimentally in a random mixture of superconducting (tin) and ferroelectric (barium titanate)  nanoparticles and the observation was analyzed in terms of epsilon near zero (ENZ) conditions (MET 2014, 2015). This paper is based on somewhat similar physics but is inspired by the recent demonstration of negative capacitance (NC) in a variety of ferroelectric materials and structures, which we believe can be exploited to induce superconductivity at relatively high temperatures. The NC state of a ferroelectric is thermodynamically unstable, and two possible approaches have been identified for accessing it as summarized in [MH 2021]. The first is to stabilize it by placing it in series with a normal insulator with a positive permittivity as originally proposed in [SSD 2008]. In this case, stability requires the overall capacitance to be positive, though individual spatial domains have been shown to be in an NC state [SSS 2019]. The second is to switch the ferroelectric from one stable state to another such that it is in a transient NC state for times $\sim$ tens of microseconds [SSA 2015]. Either the stabilized NC or the transient NC could possibly be used to demonstrate the phenomenon explored in this paper, but the latter seems more accessible experimentally, especially since transient NC is now a well-established phenomenon that has been measured by many groups using the experimental technique demonstrated in [SSA 2015].  

\begin{figure}[h!]
    \setlength\abovecaptionskip{-0.5\baselineskip}
    \centering
     \vspace{0.25in}
    \includegraphics[width=0.8\linewidth]{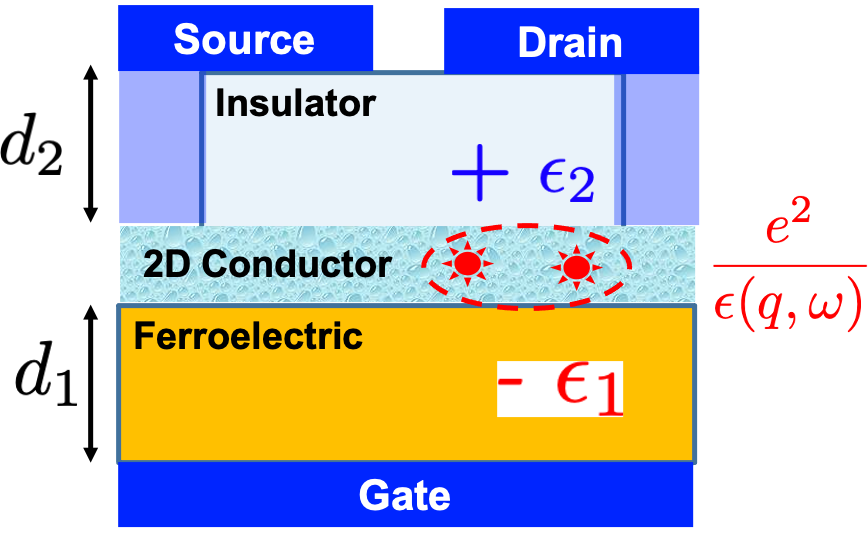} 
    \vspace{0.2in}
    \caption{\textbf{Using negative capacitance to induce superconductivity:} A two-dimensional conductor is embedded between two media, one having a negative permittivity $-\epsilon_{1}$, and another having a positive permittivity $+\epsilon_{2}$, such that the effective  permittivity $\epsilon_2 - \epsilon_1$ has a small negative value $- 2 \epsilon$. The negative capacitance state could be a temporary one attained while the ferroelectric is switching polarizations [SSA]. Both permittivities are assumed  to be \textit{isotropic}.}
    \label{fig:1}
\end{figure}

With this in mind, we consider the idealized structure shown in Fig.\ref{fig:1}a with a 2D conductor sandwiched between two media, one having a negative permittivity - $\epsilon_{1}$, and one with positive permittivity + $\epsilon_{2}$. A possible experimental structure is sketched in Fig.\ref{fig:1}b with a conducting layer on a ferroelectric insulator (see for example, FEG 2021) or a ferroelectric semiconductor recently demonstrated in [PY 2019]. The ferroelectric is assumed to be in a negative capacitance state perhaps temporarily while it is switching polarizations. In this paper the permittivities are assumed  to be \textit{isotropic}. In practice, the orientations of the ferroelectric materials should be carefully designed to maximize the effective attraction. We also note that the concept of negative capacitance goes beyond ferroelectric materials and could be engineered in magnetoelectric materials [KYC 2018] and other two-state systems separated by an energy barrier as noted in [SSA 2015].

The standard Coulomb interaction energy between two electrons in a 2D conductor embedded in a medium with a dielectric constant $\epsilon_b$ can be written as
\begin{subequations}
\begin{equation}
V(q) = \frac{e^2}{C(q)}  
\label{eq:1a}
\end{equation}
\noindent where $C(q) = 2 \epsilon_b A (q + \lambda)$, $A$ being the area of the conductor, $q$ the wavenumber and $\lambda$ the inverse 2D screening length.  For the structure  shown in Fig.\ref{fig:1}, we modify this expression to account for the two different dielectrics and their finite thicknesses:
\begin{equation}
C(q) = A \  \bigg(\overbrace{\frac{-\epsilon_1 q}{tanh(q d_1)}}^{Ferroelectric} + \overbrace{\frac{\epsilon_2 q}{tanh(q d_2)} }^{Insulator} + \overbrace{ v \frac{e^2 m}{\pi \hbar^2} F}^{2D \  conductor} \hspace{-0.09in}\bigg)
\label{eq:1b}
\end{equation}
\noindent where $m$ is the effective mass of the electrons and $v$ the number of valleys in the 2D conductor. The factor F equals one for small wavenumbers, but decreases significantly for wavenumbers $\sim 2k_f$, $k_{f}$ being the Fermi wavenumber related to the electron density $n_{s}$ by $k_{f} = \sqrt{2 \pi n_{s} / v}$ [SCR 2015]. 
\end{subequations}

It is clear from Eq.\ref{eq:1b} that with proper design the capacitance $C(q)$ can be $negative$. The interaction then becomes an \textit{attractive} one, potentially capable of inducing a transition to a superconductor-like state in the 2D conductor. We show that the corresponding transition temperature $T_{c}$ is given by a BCS-like expression
\begin{subequations}
\begin{equation}
\Delta (T \rightarrow 0) \approx \frac{min(\hbar \omega_{0}, E_{f})}{sinh(1/2 \alpha)}
\label{eq:2a}
\end{equation}
\noindent where $\Delta (T)$ is the pair potential, $\omega_{0}$ is the high frequency cut-off for the dielectric response, $E_{f}$ is the Fermi energy of the 2D conductor and $\alpha$ is the dimensionless coupling constant given by
\begin{equation}
\alpha =  \frac{e^2 m \langle \ell_{s} \rangle}{4 \pi  \epsilon_{0}  \hbar^2} 
\label{eq:2b}
\end{equation}
\noindent where $\langle \ell_{s} \rangle$ is the average value of an effective length defined in terms of the inverse negative capacitance:
\begin{equation}
\ell_{s}(k,k') \equiv  \frac{\epsilon_{0} A}{2 \pi} \int_{0}^{2 \pi} \frac{d \theta}{ - \ C(q) } \ \bigg\vert_{q = \sqrt{k^2 + k'^2 - 2k k' cos \theta}}
\label{eq:2c}
\end{equation}
\noindent $k,k'$ being wavevectors for states with energies within $\hbar \omega_{0}$ of the Fermi energy $E_{f}$. The coupling constant $\alpha$ in Eq.\ref{eq:2b} plays the role of $\textit{"N(0)V"}$ in BCS theory. Indeed it is equal to the product of the two-dimensional density of states $m/2\pi \hbar^2$ and the approximate attractive interaction energy $e^2 \langle \ell_{s} \rangle /2 \epsilon_{0}$. 

Multi-valley semiconductors introduce additional contributions from terms where $k,k'$ belong to different valleys, but in the following discussion we will assume a $\it{single \ valley}$ ($v=1$).

\end{subequations}

\section{Results}

A \textit{key point} that emerges from Eq.\ref{eq:2a} is that to observe the superconducting transition we need to design structures that maximize $\alpha$ and hence $\ell_{s}$, which in turn requires $C(q)$ to have a small negative value at least over a significant range of $\theta$. One way to achieve this is to choose $\epsilon_{1},\epsilon_{2}$ such that $\textit{for small}$ $q$ the quantity ($v$ = 1)
\begin{equation*}
- \frac{C(q \rightarrow 0)}{A}  \approx   \frac{\epsilon_{1}}{d_{1}} - \frac{\epsilon_2}{d_2} - \frac{e^2 m}{\pi \hbar^2}
\end{equation*}
\noindent has a small positive value. For example, if $ d_{1} = d_{2} = d$, we choose
\vspace{-0.1in}
\begin{subequations}
\begin{equation}
 \epsilon_{1}-\epsilon_2 \approx K \frac{e^2 m d}{\pi  \hbar^2}
\label{eq:3a}
\end{equation}
\noindent so that from Eqs.\ref{eq:2b},\ref{eq:2c}
\begin{equation}
\alpha \approx  \frac{1}{8 \pi} \int_{0}^{2 \pi} \frac{d \theta}{ K qd/tanh(qd) - F } 
\label{eq:3b}
\end{equation}
\end{subequations}
\noindent $K$ is a positive number a little larger than one, so that for small $q$ the integrand, $1/(K-1)$, is large.
\vspace{-0.2in}
\begin{figure}[h!]
    \setlength\abovecaptionskip{-0.5\baselineskip}
    \centering
     \vspace{0.25in}
    \includegraphics[width=0.75\linewidth]{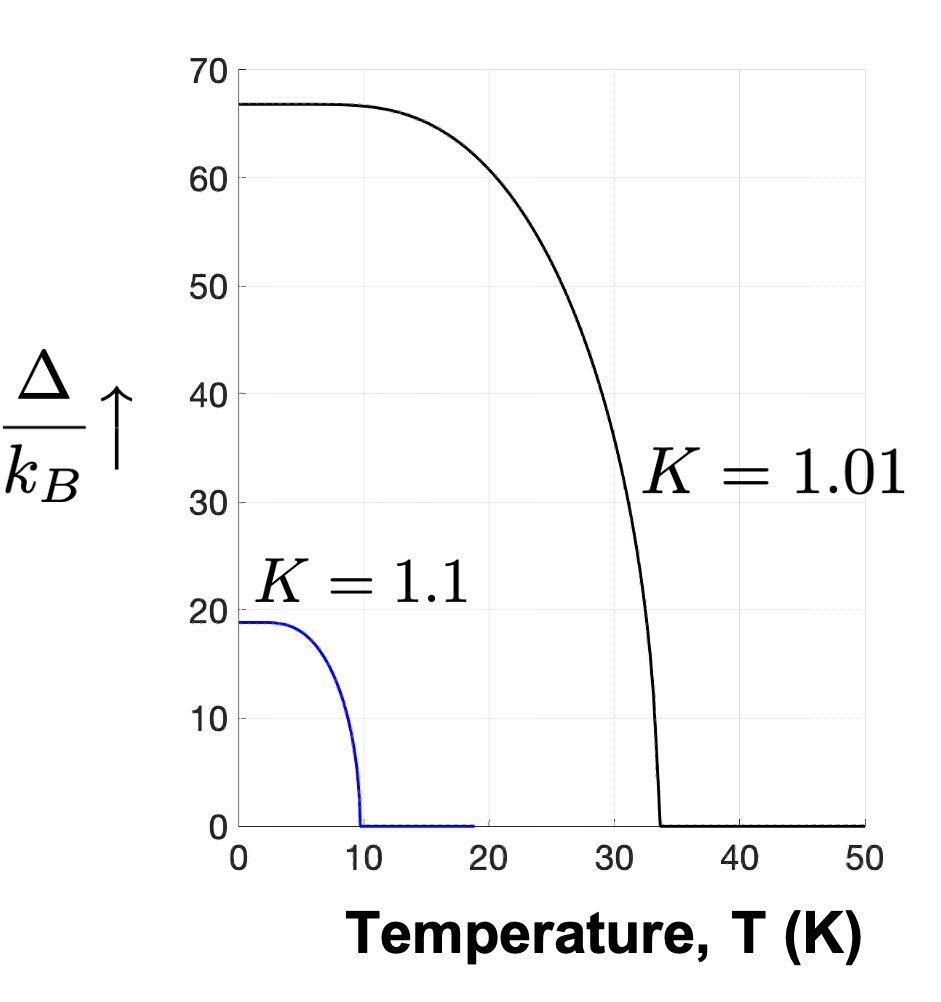} 
    \vspace{0.2in}
    \caption{\textbf{Pair potential} $\Delta (T)$ calculated from Eq.\ref{eq:6b} using two values of the constant K (see Eq.\ref{eq:3a}) as indicated. Other parameters are described in the text. The numerical values of $\Delta(T \rightarrow 0)$ agree well with Eq.\ref{eq:2a} using $\alpha$ from Eq.\ref{eq:3b}.}
    \label{fig:2}
\end{figure}

As an example, we choose $d_{1} = d_{2} \equiv d = 4 nm$ with $m=0.25 m_{0}$ and electron density, $n_s = 10^{16}/m^2$. The net negative permittivity, $\epsilon_{1}-\epsilon_{2} \approx 80 \epsilon_{0}$ as obtained from Eq.\ref{eq:3a} and the factor F is set equal to its small wavenumber value of one. We use $\omega_{0} = 2 \ THz$, comparable to the lowest atomic vibration frequency in $HfO_2$ based systems [HFO 2012]. Fig.\ref{fig:2} shows the pair potential $\Delta (T)$ calculated numerically from Eq.\ref{eq:6b} for two values of K, 1.1 and 1.01, corresponding to coupling constants $\alpha$ of $\sim0.66$ and $\sim2.13$ respectively.

\vspace{0.5in}

\section{Methods}

\subsection{Model}
\vspace{-0.1in}
Consider a two-dimensional conductor described by a parabolic band $\varepsilon_{\textbf{k}} = \hbar^2 k^2/2m$  so that the non-interacting system is described by
\begin{subequations}
\begin{equation}
H_{0} - \mu N = \sum_{\textbf{k}} (\varepsilon_{\textbf{k}} - \mu) \ c_{\textbf{k}} ^{\dagger} \  c_{\textbf{k}} 
\label{eq:4a}
\end{equation}
\noindent Here $\mu$ is the equilibrium chemical potential equal to the Fermi energy $E_{f} = \hbar^2 k_f^2/2m$.
The electron-electron interaction is described by
\begin{equation}
H_{int} =  \  \sum_{\textbf{k,k}',\beta,s',s} V(q) \ c_{\textbf{k}',s} ^{\dagger} \ c_{\beta-\textbf{k}',s'}^{\dagger} \ c_{\beta-\textbf{k,s}'}  \ c_{\textbf{k,s}}
\label{eq:4pm b}
\end{equation}
\noindent where $s,s'$ are the spin indices, $q = \vert \textbf{k-k}' \vert$ and $V(q)$ is the interaction energy from Eqs.\ref{eq:1a},\ref{eq:1b}.
\end{subequations}

\vspace{-0.1in}

\subsection{Gap equation}
\vspace{-0.1in}
It is well-established that any attractive interaction like Eq.\ref{eq:3b} can lead to a superconducting phase having a non-zero pair potential $\Delta_{k} \equiv  \langle c_{-k, \downarrow} \ c_{+k, \uparrow} \rangle$ which can be calculated by solving the \textit{BCS gap equation} self-consistently:
\begin{subequations}
\begin{equation*}
\Delta_{k}= \frac{1}{2} \ \sum_{\textbf{k'}} -V(\vert \textbf{k-k}' \vert) \ \frac{\Delta_{k'}} {E_{k'}} tanh \bigg(\frac{E_{k'}}{2k_B T}  \bigg)
\end{equation*}
\vspace{-0.3in}
\begin{equation}
\times \vartheta(\hbar \omega_0 - \vert \varepsilon_{k'} - \varepsilon_k \vert)
\label{eq:5a}
\end{equation}
\vspace{-0.2in}
\begin{equation}
where \ E_{k'} = \sqrt{(\varepsilon_{k'}-\mu)^2 + \Delta_{k'}^{2}} \hspace{1.5in}
\label{eq:5b}
\vspace{0.15in}
\end{equation}
\end{subequations}

\vspace{-0.1in}

\noindent Here we have added the last term in with the step function $\vartheta$ in Eq.\ref{eq:6a} to account approximately for the high frequency response of the dielectric response, similar to the Debye frequency used for the electron-phonon interaction in BCS theory. A full Green's function formalism could be used for a more quantitative analysis.

Defining an effective thickness
\begin{subequations}
\begin{equation}
\ell_s(k,k') \equiv \frac{\epsilon_{0} A }{2 \pi} \int_{0}^{2 \pi} \frac{d \theta}{ - \ C(q)} \ \bigg\vert_{q = \sqrt{k^2 + k'^2 - 2k k' cos \theta}}
\label{eq:6a}
\end{equation}
\noindent we can rewrite Eq.\ref{eq:4a} by substituting from Eqs.\ref{eq:1a},\ref{eq:1b} and converting the summation into an integral:
\begin{equation*}
\Delta_{k} = \frac{1}{2}  \frac{A}{4 \pi^2} \int_0^{\infty} dk' k' \ \frac{\Delta_{k'}} {E_{k'}} tanh \bigg(\frac{E_{k'}}{2k_B T}  \bigg) \hspace{0.75in}
\end{equation*}
\vspace{-0.2in}
\begin{equation*}
\times \ \frac{2 \pi e^2}{\epsilon_{0}A} \ell_s(k,k') \ \vartheta(\hbar \omega_0 - \vert \varepsilon_{k'} - \varepsilon_k \vert)
\end{equation*}
\vspace{-0.2in}
\begin{equation*}
 = \frac{e^2 m}{4 \pi \epsilon_{0} \hbar^2} \  \int_0^{\infty} d \varepsilon_{k'}\ \frac{\Delta_{k'}} {E_{k'}} tanh \bigg(\frac{E_{k'}}{2k_B T}  \bigg) \hspace{0.5in}
\end{equation*}
\vspace{-0.2in}
\begin{equation}
\times \ell_{s}(k,k') \ \vartheta(\hbar \omega_0 - \vert \varepsilon_{k'} - \varepsilon_k \vert ) \hspace{0.5in}
\label{eq:6b}
\end{equation}
\end{subequations}

\vspace{-0.15in}

\subsection{Approximate solution}
Replacing $\ell_{s}(k,k')$ with its average value, we can write Eq.\ref{eq:5a}
\begin{equation}
\Delta_{k} \approx  \underbrace{\frac{e^2 m \langle \ell_{s} \rangle}{4 \pi \epsilon_{0} \hbar^2}}_{\alpha} \ \int_{ \varepsilon_{k} - \delta}^{\varepsilon_{k} +  \delta} d \epsilon_{k'}\ \frac{\Delta_{k'}} {E_{k'}} tanh \bigg(\frac{E_{k'}}{2k_B T}  \bigg)
\label{eq:7}
\end{equation}
\noindent where $\delta = min(\hbar \omega_{0}, E_{f})$. As $T \rightarrow 0$ assuming a constant $\Delta$ in the energy range between $\mu - \delta$ and $\mu + \delta$
\begin{equation*}
\Delta \approx \alpha \ \int_{\mu - \delta}^{\mu + \delta } d \epsilon_{k'} \ \frac{\Delta} {\sqrt{(\epsilon_{k'}-\mu)^{2} + \Delta^{2}}}
\end{equation*}
\vspace{-0.2in}
\begin{equation*}
 \approx 2 \alpha \ \underbrace{\int_{0}^{\delta} \frac{\Delta \ dy} {\sqrt{y^{2} + \Delta^{2}}}}_{\Delta \ sinh^{-1}(\delta/ \Delta)} \hspace{0.5in}
\end{equation*}

\noindent This leads us to the approximate result cited earlier in Eq.\ref{eq:2a}.

\subsection{Multi-valley conductors}
Finally we note that the conduction band in many semiconductors have a number of equivalent valleys, $\it{v}$  such that $n_{s} = v \times k_f^2/2 \pi$. The gap equation is modified from Eq.\ref{eq:5a} to include a sum over the equivalent valleys. However, the intervalley terms involve large values of $q$ which require more careful discussion.

\section{Concluding remarks}
In summary, this paper draws attention to the possibility of engineering heterostructures that could provide a new route to superconductivity, based on the recent demonstration of negative capacitance, steady-state or transient. A simple treatment is presented showing that the transition temperature can be estimated from a BCS-like expression (see Eq.\ref{eq:2a}) with the coupling constant given by Eq.\ref{eq:2b} which plays the role of \textit{N(0)V} in conventional superconductivity. This expression is used to identify the key parameters that need to be optimized to observe a superconducting transition, especially the need for a small negative permittivity, which could be obtained by balancing a negative permittivity medium with a positive permittivity one. Negative capacitance  can arise in other material systems as well 
and it may be possible to engineer heterostructures with coupling constants in excess of one, leading to higher transition temperatures.

\newpage

\acknowledgements
The author is grateful to Prof. Sayeef Salahuddin for alerting him to transient NC and for helpful discussions regarding negative permittivity in ferroelectrics. He  is also indebted to Prof. Pramey Upadhyaya for his help with screening in 2D conductors.  It is a pleasure to thank Profs. Bhaskaran Muralidharan, Kerem Camsari and Zubin Jacob for their feedback on the manuscript. This work was supported by the Purdue Research Foundation.

\vspace{0.2in}

\section{References}
\begin{enumerate}

\item{\textbf{BCS 1957:} J. Bardeen, L. N. Cooper, and J. R. Schrieffer, "Theory of Superconductivity," Phys. Rev. \textbf{108} 1175 (1957).}

\item{\textbf{BLG 2018} Tommaso Cea and Francisco Guinea  "Coulomb Interaction, Phonons, and Superconductivity in Twisted Bilayer Graphene," PNAS \textbf{118} https://doi.org/10.1073/pnas.2107874118 (2018).}

\item{\textbf{CEA 2021} Tommaso Cea and Francisco Guinea  "Strong-coupling theory of condensate-mediated superconductivity in two-dimensional materials," Phys. Rev. Research, \textbf{3} 033166 (2021).}

\item{\textbf{FEG 2021:} Q. Wan, Z. Xiao, A. Kursumovic, J.L. MacManus-Driscoll and C. Durkan, "Ferrotronics for the Creation of Bandgaps in Graphene," https://arxiv.org/abs/2112.07444.}

\item{\textbf{HFO 2012:} T. J. Bright, J. I. Watjen, Z. M. Zhang, C. Muratore and A. A. Voevodin, "Optical properties of HfO2 thin films deposited by magnetron sputtering: From the visible to the far-infrared ," Thin Solid Films, \textbf{520}, 6793 (2012).}

\item{\textbf{KYC 2018:} Kerem Y. Camsari, Rafatul Faria, Orchi Hassan, Brian M. Sutton, and Supriyo Datta, "Equivalent Circuit for Magnetoelectric Read and Write Operations ," Phys. Rev. Applied, \textbf{9}, 044020 (2018).}

\item{\textbf{MET 2014:} Vera N. Smolyaninova, Bradley Yost, Kathryn Zander, M. S. Osofsky, Heungsoo Kim, Shanta Saha,
R. L. Greene and Igor I. Smolyaninov," Scientific Reports, \textbf{4},  07321 (2014).}

\item{\textbf{MET 2015:} Igor I. Smolyaninov and Vera N. Smolyaninova," Metamaterial Superconductors," Phys. Rev. B, \textbf{91},  094501 (2015).}

\item{\textbf{MH 2021:} Michael Hoffmann, Stefan Slesazeck and Thomas Mikolajick, "Progress and future prospects of negative capacitance electronics: A materials perspective," APL Materials \textbf{9} 020902 (2021).}

\item{\textbf{MLC 1964:} M.L.Cohen, "Superconductivity in many-valley semiconductors and semimetals," Phys. Rev. \textbf{134} A511 (1964).}

\item{\textbf{PBA 1982:} Philip B. Allen and Bozidar Mitrovic, "Theory of Superconducting $T_c$," Solid State Physics \textbf{37}, 1 (1982).}

\item{\textbf{PY 2019:} Mengwei Si, Atanu K. Saha, Shengjie Gao, Gang Qiu, Jingkai Qin, Yuqin Duan, Jie Jian, Chang Niu, Haiyan Wang, Wenzhuo Wu, Sumeet K. Gupta, and Peide D. Ye, "A Ferroelectric Semiconductor Field-Effect Transistor," Nature Electronics \textbf{2}, 580 (2019).}

\item{\textbf{PWA 1958:} P.W.Anderson, "Random Phase Approximation in the Theory of Superconductivity" Phys. Rev. \textbf{112} 1900 (1958).}

\item{\textbf{SAI 2021} See for example, Yu Saito, Tsutomu Nojima and Yoshihiro Iwasa, "Highly Crystalline 2D Superconductors," Nature Reviews Materials, \textbf{2} 16094 (2021).}

\item{\textbf{SCR 2015} S. Das Sarma and E.H. Hwang, "Screening and Transport in 2D Semiconductor Systems at Low Temperatures," Scientific Reports, \textbf{5} 16655 (2015).}

%\item{\textbf{SCR 2009:}  T. Ando, M. M. Frank, K. Choi, C. Choi, J. Bruley, M. Hopstaken, M. Copel, E. Cartier, A. Kerber, A. Callegari, D. Lacey, S. Brown, Q. Yang, V. Narayanan, "Understanding mobility mechanisms in extremely scaled HfO2 (EOT 0.42 nm) using remote interfacial layer scavenging technique and Vt-tuning dipoles with gate-first process," 2009 IEEE International Electron Devices Meeting (IEDM), https://ieeexplore.ieee.org/document/5424335.}

\item{\textbf{SSA 2015:} Asif Islam Khan, Korok Chatterjee, Brian Wang, Steven Drapcho, Long You, Claudy Serrao, Saidur Rahman Bakaul, Ramamoorthy Ramesh and Sayeef Salahuddin, "Negative capacitance in a ferroelectric capacitor," Nature Materials \textbf{14}, 182 (2015).}

\item{\textbf{SSD 2008} S. Salahuddin and S.Datta, "Use of Negative Capacitance to Provide Voltage Amplification for Low Power Nanoscale Devices", NanoLetters \textbf{8}, 405 (2008).}

\item{\textbf{SSS 2019:} Ajay K Yadav, Kayla X Nguyen, Zijian Hong, Pablo García-Fernández, Pablo Aguado-Puente, Christopher T Nelson, Sujit Das, Bhagwati Prasad, Daewoong Kwon, Suraj Cheema, Asif I Khan, Chenming Hu, Jorge Íñiguez, Javier Junquera, Long-Qing Chen, David A Muller, Ramamoorthy Ramesh, Sayeef Salahuddin, "Spatially resolved steady-state negative capacitance," Nature \textbf{565}, 468 (2019).}

%\item{\textbf{TDG 2011:} Rafi Bistritzer and Allan H. MacDonald  "Moire Bands in Twisted Double-layer Graphene," PNAS \textbf{108} 12233 (2011).}

\end{enumerate}

%\item{\textbf{NC:} Eugene A. Eliseev, Mykola E. Yelisieiev, Sergei V. Kalinin and Anna N. Morozovska, "Whither Steady-state Negative Capacitance of a Ferroelectric Film ," https://arxiv.org/abs/2112.04712.}

%\section{Extras}
%\begin{enumerate}
%\item{\textbf{ABB:} D. Allender, J. Bray and J. Bardeen, "Model for an Exciton Mechanism of Superconductivity," Phys. Rev. \textbf{B7}, 1020 (1973).}
%\end{enumerate}

%Lindhard function defined as
%\begin{equation}
%F(\beta) = \frac{1}{2} + \frac{1-x^2}{4x} log \bigg\vert \frac{1+x}{1-x} \bigg\vert \ , \ x \equiv \frac{\beta}{2k_f}
%\label{eq:1c}
%\end{equation}

% \bibliographystyle{naturemag}
% \balance\bibliography{library}
\end{document}